\documentclass[twocolumn,showpacs,showkeys,preprintnumbers,amsmath,amssymb]{revtex4}
\usepackage{graphicx}% Include figure files
\usepackage{dcolumn}% Align table columns on decimal point
\usepackage{bm}% bold math
\begin{document}
\title{Empirical Estimates of the Neutron-Nucleus Scattering Cross Sections}
\author{S.S.V.   Surya  Narayan$^\dagger$\footnote  {\normalsize{  \it  The
author Surya Narayan's name  appeared also  as "  S.V.S.  Sastry  "  in
Nuclear  physics  journals.  }},
Rajesh S. Gowda$^+$ and S. Ganesan$^+$}
\affiliation{  $^\dagger$  Nuclear  Physics Division, $^+$ Reactor Physics Design
Division,\\
Bhabha Atomic Research Centre, Trombay, Mumbai 400 085, India}
%\email{snarayan@apsara.barc.ernet.in}
\begin{abstract}{
Theoretical  study  of  systematics of neutron scattering cross sections on
various materials for neutron energies up to several  hundred  MeV  are  of
practical  importance. In this paper, we analysed various cross sections of
neutron-nucleus (n-N) scattering for several systems in the energy range of
50-250  MeV,  predicted  by  the  optical  model   using   Koning-Delaroche
potentials.  We  propose  an empirical approach to successfully predict the
energy dependence of total, shape elastic, reaction cross sections and zero
degree scattering angular distributions. We demonstrate that owing  to  two
conditions,  only  two  out  of these four cross sections need to be fitted
empirically. Further, we modified the Wick's  approximation  for  the  zero
degree  angular  distributions and use this approach for estimating various
cross sections for any nucleus from Aluminum to Lead.
}\end{abstract}
\date  {\today}
\pacs{ 24.10.Ht, 25.40.-h, 28.20.Cz}
\keywords{Optical  Model,  total  cross  section,  reaction, shape elastic,
forward elastic cross sections, n-N  scattering,  Wick's  limit,  empirical
formulae.}
\maketitle
\par
\noindent
In  recent  times,  the concept of an accelerator driven sub-critical (ADS)
system is drawing   worldwide  attention  \cite{ads1,ads2}.  In  this   ADS
system,  neutrons  are produced by bombarding a heavy element target with a
high energy proton beam of typically above 1.0GeV with a current of $>10mA$
\cite{ads1}. Such a system serves a dual purpose of  energy  multiplication
and  waste  incineration. In this context, theoretical study of systematics
of neutron scattering cross  sections  on  various  materials  for  neutron
energies up to  several  hundred  MeV  are of practical importance. In this
paper, we propose an empirical approach to successfully predict the  energy
dependence  of  total  cross sections ($\sigma_{tot}$), shape elastic cross
sections ($\sigma_{se}$), reaction  cross  sections  ($\sigma_{reac}$)  and
zero    degree    scattering    angular    distributions    ($\sigma_0$~or~
$\sigma(\theta=0$)), of the neutron-nucleus  (n-N)  scattering.  The  total
cross  sections are usually fitted by using the Ramsauer model. The nuclear
Ramsauer model was first proposed by Lawson in the year 1953  as  a  simple
means  to  understand  the  energy  dependence  of  total cross sections of
neutron nucleus scattering. In  order  to  appreciate  this  model,  it  is
necessary  to discuss briefly the optical model (OM) description of neutron
scattering. In the OM approach, complex optical model potentials (OMP)  are
used  and  the  Schrodinger's  equation  is solved to obtain the scattering
amplitude. The real part of  the  OMP  describes  the  scattering  and  the
imaginary  part  results in attenuation or absorption of the incident wave.
This absorption gives an estimate  of  the  optical  model  reaction  cross
section.   The  calculations  are  usually  performed  using  partial  wave
expansion   method   and   the   phase    shifts    ($\eta_\ell=\alpha_\ell
e^{i\beta_\ell})$  are  determined.  These complex phase sifts are strongly
angular momentum and energy dependent for a given  set  of  potentials.  In
terms  of  the  phase  shifts  and  the  wave  number ($ \lambdabar = \hbar
/\sqrt{2mE}$), various cross sections are given below.
\begin{eqnarray}
\sigma_{tot}&=&2\pi\lambdabar^2\sum_\ell(2\ell+1)\left[1-\Re{\eta_\ell }\right] \label{st-om} \\
\sigma_{se}&=&\pi\lambdabar^2\sum_\ell(2\ell +1)|1-\eta_\ell |^2  \label{se-om} \\
\sigma_{reac}&=&\pi\lambdabar^2 \sum_\ell(2\ell +1)\left(1-|\eta_\ell |^2\right)  \label{r-om} \\
\frac{d\sigma}{d\Omega}(\theta)&=&\frac{\lambdabar ^2}{4} \left|\sum_\ell (2\ell +1)(1-\eta_\ell )P_\ell (\cos{\theta})\right|^2  \label{s0-om}
\end{eqnarray}
\par
\noindent
Extensive  study  of the optical model fits of scattering cross sections on
various nuclei over wide energy range have been  made  by  several  groups.
This  is  owing  to  the  excellent  data  base  of  neutron cross sections
available in the energy range up to 600 MeV  \cite{finlay,dietrich1,abfal}.
The  most  recent  work  by Koning and Delarosche \cite{kd} presents a very
exhaustive search for OMP parameters that fit the data very well up to  200
MeV.  In  the  remaining  part  of  our  work,  we  use  the  OM code SCAT2
\cite{scat2} with Koning and Delaroche potentials. We  treat  this  as  the
"experimental  data"  for  shape  elastic, reaction, total scattering cross
sections with spherical targets. We made a phenomenological Ramsauer  model
analysis  of  this  data to derive systematics of Ramsauer model parameters
and utilize it for further predictions. As mentioned  before,  the  nuclear
Ramsauer  model  \cite{lawson}  provides a simple means to parameterise the
energy dependence of neutron nucleus total scattering cross sections.  This
model  assumes  that the scattering phase shifts are independent of angular
momentum ($\ell$) as  given  in  Eq.(5)  ($\eta=\alpha  e^{i\beta})$  ,  in
contrast  to the predictions of the optical model given in Eq.(1). Further,
it was proposed that the $\ell$-independent phase shift varies slowly  with
energy. The model thus implies that the nucleus is seen as a right circular
cylinder by the neutrons incident along the axis of symmetry. Owing to this
unphysical  picture,  initially  this model did not receive much attention,
despite a successful demonstration of this  model  for  neutron  scattering
from  various  nuclei  by  Peterson  \cite{peterson,book}.  There were some
attempts  to  put  this  Ramsauer  model  on  a  sound  theoretical   basis
\cite{franco,gould,anderson,grimes1}  (see references therein). The neutron
total cross sections have thus been well studied using this model,  over  a
wide  range  of  nuclear  masses  as well as neutron energies up to 500 MeV
\cite{anderson,bauer,madsen,grimes1,grimes2,grimes3,dietrich2}.     Various
cross sections used in the Ramsauer model are given below.
\begin{eqnarray}
\sigma_{tot} &=& 2\pi\ (R+\lambdabar)^2\left(1-\alpha\cos{\beta}\right) \label{eqst} \\
\sigma_{se} &=& \pi\ (R+\lambdabar)^2\left[ \left((1-\alpha\cos{\beta}\right)^2 + \left(\alpha\sin{\beta}\right)^2 \right]  \label{eqse}
\end{eqnarray}
\noindent
\section{Analysis of total cross sections}
\noindent
We  performed  the  Ramsauer  model  fits  to total cross sections by using
Eq.(5,7) for a set of seven nuclei with  a  $  \chi^2/N$    =  653.762  .  The
$R,\alpha,\beta$ are functions of atomic mass number and the center of mass
energy.
\begin{eqnarray}
\sigma_{tot} &=& 2\pi\ (R+\lambdabar)^2\left(1-\alpha\cos{\beta}\right) \label{eqstfit}\\
\alpha &=& \alpha_0/A^{1/3} ~;~ \beta/A^p=a_0   \left(\sqrt{b_0+c_0E}-\sqrt{E}\right) \label{stalbta}\\
r_0&=&1.25519, ~~    \alpha_0=1.42042,~p = 0.29361 \\
~a_0&=&3.0,  ~b_0=5.84633, ~ c_0=0.97676 \nonumber \label{stfitpar}
\end{eqnarray}
The  $\sigma_{tot}$  cross  sections  are  shown in Fig. 1 with solid lines
representing Ramsauer model fits using Eqs.(7-9) and the symbols  represent
the  results  from  the optical model code SCAT2 \cite{scat2}. The fits are
obtained with total of six free parameters as given in  Eq.(9),  over  wide
mass  range of $^{27}Al$ to $^{208}Pb$ and also covering the neutron energy
region ($E_{cm}$) of 50-250 MeV. Similar Ramsauer model fits to total cross
sections were already shown by various groups  \cite{grimes1,grimes2}  (and
see  the  references  therein). In these works, the normalised (relative to
$^{208}Pb$) cross sections were shown for  various  nuclei  \cite{grimes2}.
Our  functional  dependence  on  energy  and  mass given in Eq.(8) with six
global parameters was able to reproduce the SCAT2 results  very  well.  The
form of $\beta$ given in Eq.(8) is chosen such that $\beta/A^p$ is a global
phase  function for all nuclei, as shown versus energy in Figure 2. We have
chosen this form of $\alpha,\beta$ which directly scale by  some  power  of
mass  rather than a polynomial function of mass in order to obtain a global
function. The energy dependence of global $\beta$ function  has  also  been
well discussed in literature. The parameter $b_0$ is called potential depth
and  $c_0$  ($\approx  0.90$) is the non local parameter. The atomic number
dependence of $\beta$ can also  be  introduced  through  these  parameters,
however at the cost of global function.
\begin{figure}
\includegraphics[width=8.0cm,height=9.0cm]{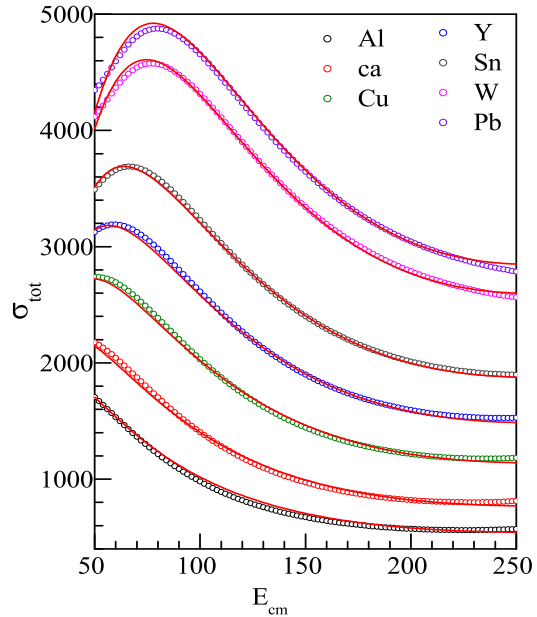}
\caption{Ramsauer  model  fits  (solid  lines)  to  total  scattering cross
sections of SCAT2  (symbols)  versus  E$_{cm}$,  using  Eq.  (7).  The  six
parameters required are mentioned in the text in Eqs.(8,9). The curves from
bottom to top are respectively for Al,Ca,Cu,Y,Sn,W,Pb.}
\label{sigtot}
\end{figure}
\begin{figure}
\includegraphics[width=8.0cm,height=7.0cm]{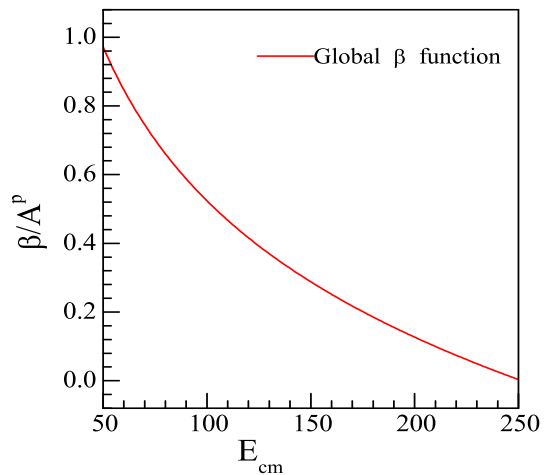}
\caption{The  global  $\beta$  function required for Ramsauer model fits to
total scattering cross sections using Eqs. (8,9)}.
\label{globalfn}
\end{figure}
\section{Relations between various cross sections }
The  total  scattering  cross  section is defined as a sum of shape elastic
and reaction cross sections given in  Eq.(10).
\begin{equation}
\sigma_{tot}=\sigma_{se} + \sigma_{reac}
\label{reltot}
\end{equation}
Further,  it  has been observed \cite{gama01,gama02} earlier that the ratio
(see Eq.(11)) of zero degree elastic cross section and  the  shape  elastic
cross  sections  shows linearity with energy. In the present work, we study
this linearity in the energy range of 50-250  MeV.  We  observed  that  the
linear  relation  is  quite  good at higher energy region above 150 MeV and
this linearity remains valid over a large energy range  at  high  energies.
These  ratios from SCAT2 for various nuclei are shown in Fig. 3 by symbols.
They have been parameterised well (see solid lines in figure) by a function
of the form given by Eq.(12). Making use of  the  Ramsauer  assumption  for
these  cross  sections  in  Eqs.(2,4), it can be shown that the ratio has a
simple  form  as  in  Eq.(12).  This  is  because,   the   ratio   involves
$(\ell_{max}+1)$    where   we   use   $\ell_{max}=\frac{R}{\lambdabar}   +
\frac{1}{2}$. For $\gamma_0$ fits we used $R=1.254A^{1/3}$fm in Eq.(12).
\begin{eqnarray}
\gamma_0 &=&\frac{ \frac{d\sigma}{d\Omega}(\theta=0)}{\sigma_{se}} \label{gama0def}\\
\gamma_0 &=& \left(\frac{R}{\lambdabar}+1.5\right)^2/{4\pi} \label{gama0fit}
\end{eqnarray}
Therefore,  inverting this equation and using the known shape elastic cross
sections, we can obtain the zero degree elastic cross sections for a  given
energy.  Therefore,  $\gamma_0$  serves as a relation between $\sigma_{se}$
and $\sigma(\theta=0)$.
\begin{equation}
\frac{d\sigma}{d\Omega}(\theta=0) =\gamma_0 \sigma_{se}
\end{equation}
\begin{figure}
\includegraphics[width=8.0cm,height=8.5cm]{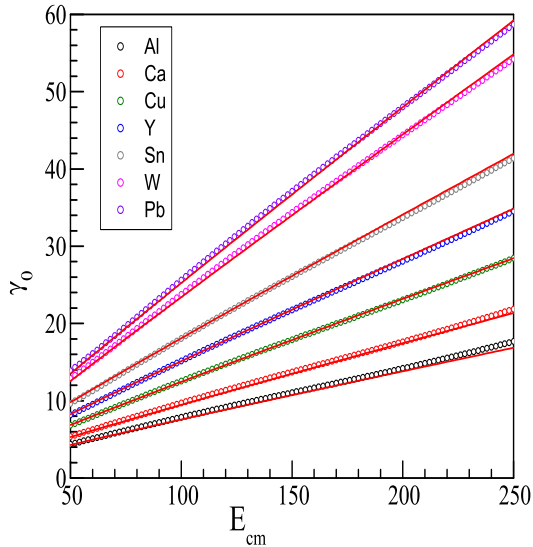}
\caption{The  $\gamma_0$  factor  defined  (Eq.(11))  as  the ratio of zero
degree angular distributions and shape elastic distributions. This  can  be
fitted by a functional form given in Eq.(12).}
\label{gama0fig}
\end{figure}
\section{Analysis of Reaction Cross Sections}
As  mentioned  in  the introduction, we need to fit only two quantities and
the other  two  will  be  derived.  In  section  I,  we  parameterised  the
$\sigma_{tot}(A,E)$  for  all  nuclei.  The two important relations between
various cross  sections  have  been  discussed  in  the  previous  section.
Therefore,  it  suffices  to  parameterise one of the three remaining cross
sections $\sigma_{reac},\sigma(\theta=0),\sigma_{se}$. In this  section  we
analyse  the  $\sigma_{reac}$,  the  reaction  cross  sections  and  obtain
$\sigma(\theta=0),\sigma_{se}$ as derived quantities. As shown  in  Eq.(3),
the  reaction  cross sections involves only the transmission function {$\it
i.e.,$} (1-$|\eta_\ell|^2$). Unlike the $\sigma_{tot},\sigma_{se}$ and  the
angular distributions,~~ the reaction cross sections do not involve $\beta$
phase  function.  Therefore, we proceed to obtain the empirical fits to the
reaction cross sections predicted by SCAT2 code. For this fits, we used the
functional form and parameters given below, yielding a $\chi^2/N$=323.641
\begin{eqnarray}
\sigma_{reac}&=&\pi\left(R+1.5\lambdabar\right)^2 e^{-\alpha}~~;~~\alpha=\alpha_0 \sqrt{E}/A^p \label{eqsrfit}\\
\alpha_0&=&0.20873~;~r_0=1.38413~;~p=0.42507  \label{srfitpar}
\end{eqnarray}
It  should  be noted that the fitted parameters of Eq.(15) may not be equal
to  the  corresponding  parameters  of  Eq.(9),  as  these  are   effective
parameters  representing  different  phenomenological functional forms. One
should also note that  the  multiplying  radius  dependence  of  these  two
equations   are   respectively,   $2\pi\left(R+\lambdabar\right)^2   $  and
$\pi\left(R+1.5\lambdabar\right)^2$. These $\sigma_{reac}$ fits using three
parameters given in Eqs.(14,15), are shown in  Fig. 4  for all  the  nuclei
studied.  Figures  (1,4)  show that these cross sections of SCAT2 (symbols)
are well reproduced by our empirical fitting  functional  forms  and  their
parameters   (lines  in  figures).  Hence,  we  proceed  to  utilise  these
$\sigma_{reac}$ fits to obtain other  cross  sections  predicted  by  SCAT2
code.
\begin{figure}
\includegraphics[width=8.0cm,height=8.5cm]{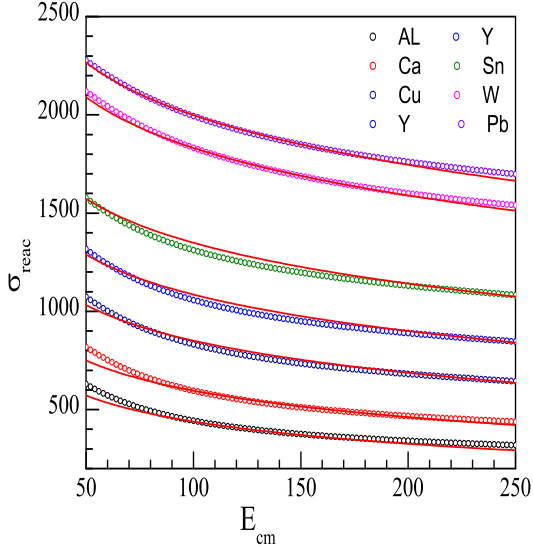}
\caption{Phenomenological  fits  (lines)  to  reaction  cross  sections  of
optical model code SCAT2. The SCAT2 data (symbols) is fitted by exponential
energy dependence multiplied by $(R+1.5\lambdabar)^2$ (see text). The  fits
need three parameters that are given in text. The curves from bottom to top
are respectively for Al,Ca,Cu,Y,Sn,W,Pb.}
\label{sigr}
\end{figure}
\begin{figure}
\includegraphics[width=8.0cm,height=12.0cm]{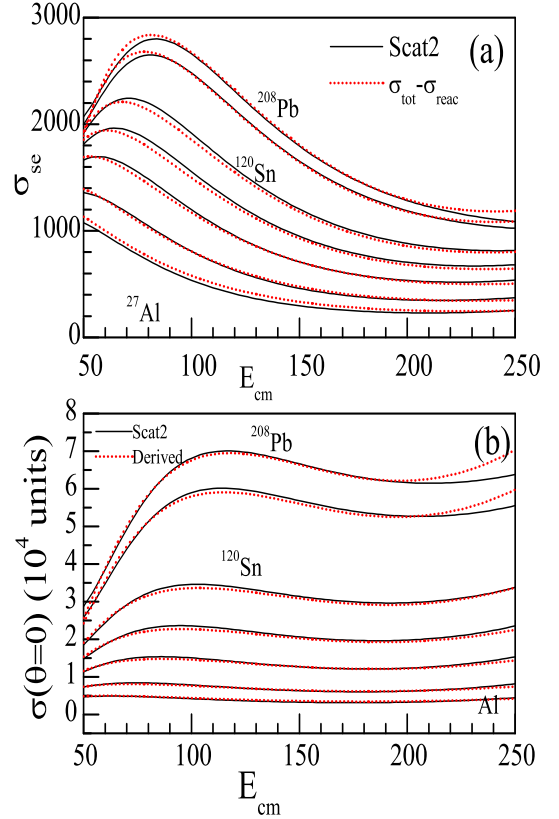}
\caption{The  derived  cross  sections  for  shape elastic (a) and the zero
degree elastic angular distributions (b). The shape elastic is obtained  as
difference  of  total  and  reaction  cross sections. The zero degree cross
sections are obtained by using $\gamma_0$ factor form shape elastic data.}
\label{rses0}
\end{figure}
\noindent
The  difference of the two parameterised cross sections gives shape elastic
scattering as given by,  $\sigma_{se}$=$\sigma_{tot}  -  \sigma_{reac}$  as
shown  in  Fig. 5(a). Making use of the $\gamma_0$ and the $\sigma_{se}$ of
Fig. 5(a), the angular  distribution  at  zero  degree  versus  energy  are
obtained as shown in Fig. 5(b). In summary, we have fitted total scattering
cross  sections  and  reaction  excitation  functions  and hence we derived
$\sigma_{se},\sigma_{elas}(\theta=0)$. These  two  derived  cross  sections
also  agree very well with the SCAT2 code predictions. \section{Wicks Limit
Modification for Forward Elastic Scattering} In this section,  we  describe
the   parameterization   of   energy  dependence  of  zero  degree  angular
distributions and hence the  $\sigma_{reac},\sigma_{se}$  will  be  derived
quantities.  It  is  well  known that the optical theorem relates the total
scattering cross section  to  the  imaginary  part  of  forward  scattering
amplitude  is  given  by  Eq.  (16).  However,  the forward elastic angular
distributions are defined in Eq. (17). In the Wick's limit \cite{wick}, the
real part of the forward scattering amplitude is neglected as in Eq.  (18).
Therefore, the fractional deviation ($\delta$) of true angular distribution
at  zero  degree for a given energy from the Wick's limit is defined in Eq.
(19).
\begin{eqnarray}
\sigma_{tot}&=& \frac{4\pi}{k} \Im{f(0^o)}  \label{opth} \\
\sigma_{0} &=& |f(\theta=0)|^2 = \left(\Im{f(0)}\right)^2 + \left( \Re{f(0)}\right)^2  \label{sig0} \\
\sigma_{0} &\approx& \sigma_{0}^W =  \left( \Im{f(0)}\right)^2 =\left(\frac{k}{4\pi}\sigma_{tot}\right)^2.  \label{wsig0} \\
\delta &=& \frac{\left(\sigma_{0}-\sigma_0^W\right)}{\sigma_0^W}  \label{eqdelta}
\end{eqnarray}
Whenever,  $\delta$  is  small,  Wick's limit gives a good approximation to
zero degree  angular  distributions.  In  depth  study  of  this  deviation
function  has  been performed for several nuclei by Dietrich {\it et. al.,}
\cite{dietrich2}. They have presented the deviation function  derived  from
optical  model  predictions  and  Wick's limit values. In the present work,
following \cite{dietrich2}, we have constructed the deviation function over
50-250 MeV center of mass energy range for seven target nuclei. We observed
that these can be parameterised in  terms  of  a  Woods-Saxon  function  of
energy, whose parameters are nuclear mass number dependent. Given below are
the functional form for the deviation function and its parameters optimised
for all nuclei.
\begin{eqnarray}
\delta &=& \frac{\left(V_1 + V_2E\right)}{1+e^{(E_0-E)/a_0} }  \\
E_0 &=& 61.557 + 0.47825 A -9.90931~ 10^{-4}A^2  \\
V_2 &=& 0.00217-2.26414~10^{-5}A +5.164~10^{-8}A^2  \nonumber \\
V_1 &=& 0.15 + 143.04/A^2  \nonumber \\
a_0 &=& 13.48384 \nonumber
\end{eqnarray}
The  deviation function derived from SCAT2 and its empirical fits are shown
in Fig. 6. As shown in figure,  the  function  with  nine  free  parameters
listed  above,  gives very good fits to data of SCAT2. We have also tried a
logarithmic function, which gives similar  quality  of  fits  to  deviation
function  with fewer number of parameters. It should be noted here that the
$\delta$ function can be given  by  some  empirical  prescription  for  all
nuclei,  very  much  similar  to  the  fits  of total cross sections. Once,
$\delta$ of Eq.(19) is known,  using  $\sigma_0^W$  of  Eq.(18),  the  zero
degree elastic angular distribution versus energy can be obtained, as given
by $\sigma_0 = (1+\delta) \sigma_0^W$. Consequent to knowing $\sigma_0$, we
can  use  $\gamma_0$  of  Eq.(11)  to  get the shape elastic cross sections
$\sigma_{se}$. By subtracting the $\sigma_{se}$ from total  cross  section,
we  can  fix  the  reaction  cross  section versus energy. We followed this
scheme and the results are  shown  in  Fig. 7. The  agreement  between  the
derived  quantities  from  fits  and the SCAT2 predictions is very good. It
should be noted that the deviation function method accomplishes fitting  of
the  zero  degree  angular  distribution  without  need for a phase $\beta$
function. Otherwise, the angular distributions  fits  would  need  a  phase
function.
\begin{figure}
\includegraphics[width=8.0cm,height=8.5cm]{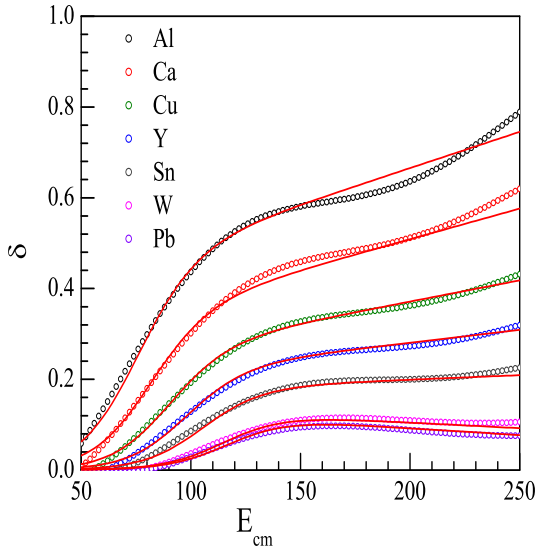}
\caption{The  empirical  fits  of  deviation  function  versus  energy,  as
explained in the text.}
\label{eta}
\end{figure}
\begin{figure}
\includegraphics[width=8.0cm,height=12.0cm]{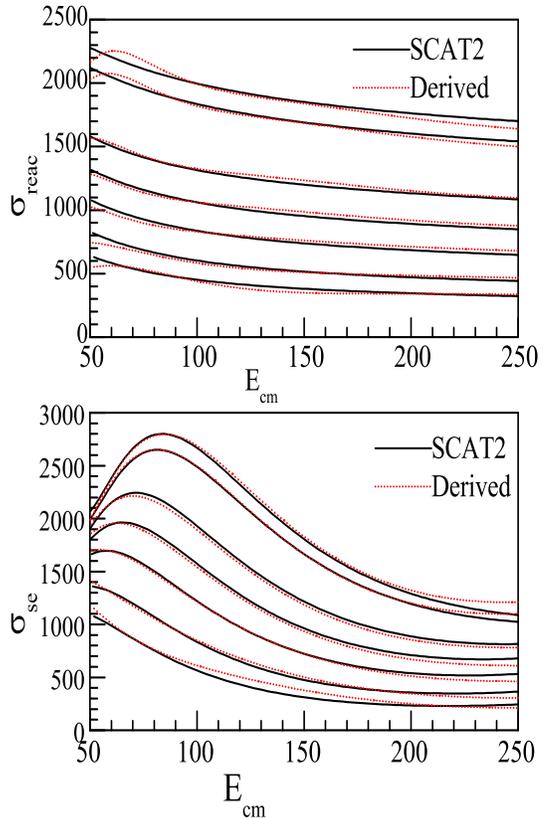}
\caption{The  derived  cross  sections  for  shape elastic and the reaction
cross sections.}
\label{wick-all}
\end{figure}
\section{Analysis of Shape Elastic cross sections}
In  this  section,  we show the parameterization of the shape elastic cross
sections, which need a phase  $\beta$  function.  The  parameterization  is
performed  very  much  similar  to  the  fits  of  total cross sections. We
followed Eq. (6) for the functional form and searched for  the  parameters.
The  $R,\alpha,\beta$ are functions of atomic mass number and the center of
mass energy. The best fit functional form  and  the  parameters  are  given
below   which   yielded   $   \chi^2/N$   =   1222.97   However,  in  these
parameterizations,  the  $\beta$  function  is  more   complicated,   while
maintaining the functional forms for other parameters.
\begin{eqnarray}
\sigma_{se} &=&F\left[\left(1-\alpha\cos{\beta}\right)^2
+ \left(\alpha \sin{(\beta\alpha_2)} \right)^2 \right]  \label{eqsefit}\\
F &=&\pi\ (R+1.5\lambdabar)^2 ,~r_0=1.20815  \\
\alpha &=& \alpha_0/A^{1/3} ~~; \alpha_{2} = \alpha_{20}/A^{2/3} \\
\beta&=& A^p a_0   \left(\sqrt{b_0+c_0E}-\sqrt{E}\right) \\
\alpha_0&=&1.08627,~ \alpha_{20}=12.70632,~ p = 0.3146 \\
a_0&=&0.5534,  ~b_0=33.0085, ~ c_0=0.8626 \nonumber \label{sefitpar}
\end{eqnarray}
The  shape  elastic  cross  sections  fits  are shown in Fig. 8 for all the
systems, and the quality of fits  is  good.  Following  the  shape  elastic
parameterization,  the  forward  angular distribution can be obtained using
$\gamma_0$ factor. The reaction is given as difference of total  and  shape
elastic     cross     sections.     The     two     derived     quantities,
$\sigma_{reac},\sigma_{0}$ agree well with SCAT2 data (not shown here).  In
summary,  we considered three methods in sections III,IV,V where two of the
four cross sections were parameterised and the remaining two were  derived,
as shown in the table.
\begin{figure}
\includegraphics[width=8.0cm,height=8.5cm]{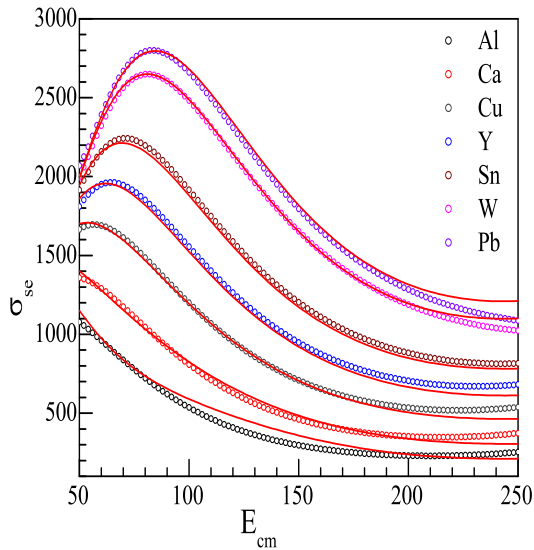}
\caption{Empirical  fits  of  shape  elastic  cross  sections.  The fitting
functions and parameters are given in Eqs.(22-26)}
\label{sefit}
\end{figure}
%%%------------table of ------------------%%%%%%%%
\begin {center}{
\begin{table}
\begin{tabular}
{|c|c|c|}
\hline
Method &  Fitted quantities          &   Derived quantities   \\
\hline
1.     & $\sigma_{tot},\sigma_{reac}$     &  $\sigma_{se},\sigma_0$ \\
\hline
2.    & $\sigma_{tot},\sigma_0$     &  $\sigma_{se},\sigma_{reac}$ \\
\hline
3.    & $\sigma_{tot},\sigma_{se}$  &  $\sigma_{reac}, \sigma_0$\\
\hline
\end {tabular}
\caption{List of parameterised cross sections and the derived quantities in
sections  of  the  text. In above $\sigma_0$ represents zero degree angular
distributions.}
\end{table}
}\end {center}
\par
\noindent
{\bf Conclusion}\\
\par
\noindent
We  have  analysed  the predictions of optical model code SCAT2 for neutron
nucleus scattering  using  Koning  Delaroche  potentials.  We  perform  the
Ramsauer  model  parameterization  of  total  scattering cross sections and
derive the systematics of the Ramsauer  parameters.  We  have  proposed  an
empirical  fits to the reaction cross sections of SCAT2 and using this, the
shape elastic cross sections have been obtained as the difference of  these
two  empirical  results.  Using the $\gamma_0$ factor which is the ratio of
forward elastic scattering to the shape elastic  scattering,  we  determine
the  forward  elastic  angular  distributions. Thus our empirical method is
able to predict the SCAT2 results with a good accuracy for all  these  four
quantities,                                                          namely
$\sigma_{tot},\sigma_{reac},\sigma_{se},\sigma(\theta=0)$. We have  studied
the  Wick's  limit  of  the  forward  elastic cross sections from SCAT2. We
parameterised the deviation function of these  angular  distributions  from
Wick's  limit  and  using  these as a alternative method, all the remaining
cross sections have been obtained which compare well  with  SCAT2  results.
The  present  method  is  about  empirical  estimates  of the various cross
sections, therefore it will be of practical importance wherever high energy
neutron  scattering  cross  sections  are  required  for  example  for  the
accelerator driven sub critical assemblies.\\
\medskip
\par
\noindent
We  acknowledge  fruitful  discussions  with  Dr.  S.  Kailas  and Dr. A.K.
Mohanty. We acknowledge Mr. Mukadam  Mayuresh  for  technical  support.
\par
%%------------------------------------------------

%%%%%%--------------------------------------------------------------------
\end{document}